\input psfig.sty
\def\vcr{{V_{\rm cr}}}
\def\Xp{X^{\prime}}
\def\Yp{Y^{\prime}}
\def\pa{\partial}
\def\o{\over}

\def\pn{{\par\noindent}}

\font\small = cmr7
\def\thn{{\thinspace}}
\def\sc{\scriptstyle}
\def\Msun{\hbox{$\thn M_{\odot}$}}

\def\={\thn\thn=\thn\thn}
\def\halfline{{\null \vskip 7 truept}}
\def\tgs{{\thn \rlap{\raise 0.5ex\hbox{$\sc  {>}$}}{\lower 0.3ex\hbox{$\sc  {\sim}$}} \thn }}
\def\tls{{\thn \rlap{\raise 0.5ex\hbox{$\sc  {<}$}}{\lower 0.3ex\hbox{$\sc  {\sim}$}} \thn }}
\def\tll{{\raise 0.3ex\hbox{$\sc  {\thn \ll \thn }$}}}
\def\tgg{{\raise 0.3ex\hbox{$\sc  {\thn \gg \thn }$}}}
\def\tle{{\raise 0.3ex\hbox{$\sc  {\thn \le \thn }$}}}
\def\tge{{\raise 0.3ex\hbox{$\sc  {\thn \ge \thn }$}}}
\def\tl{{\raise 0.3ex\hbox{$\sc  {\thn < \thn }$}}}
\def\tg{{\raise 0.3ex\hbox{$\sc  {\thn > \thn }$}}}
\def\ts{{\raise 0.3ex\hbox{$\sc  {\thn \sim \thn }$}}}

\def\do{{\tt "}}
\def\tp{{\raise 0.3ex\hbox{\small +}}}

\def\eq{{\thn\equiv\thn}}

\def\w{{\ \ \ \ \ }}
\def\sep{{\par \noindent \hangindent=15pt \hangafter=1}}
\def\deg{{^\circ}}

\def\z{\ \ \ \ }

\def\asec{{^{\prime\prime}}}
\def\hp{H_p}
%
%
%

\ifx\mnmacrosloaded\undefined \input mn\fi

%

\newif\ifAMStwofonts

\ifCUPmtplainloaded \else
  \NewTextAlphabet{textbfit} {cmbxti10} {}
  \NewTextAlphabet{textbfss} {cmssbx10} {}
  \NewMathAlphabet{mathbfit} {cmbxti10} {} 
  \NewMathAlphabet{mathbfss} {cmssbx10} {} 
  \ifAMStwofonts
    \NewSymbolFont{upmath} {eurm10}
    \NewSymbolFont{AMSa} {msam10}
    \NewMathSymbol{\upi}     {0}{upmath}{19}
    \NewMathSymbol{\umu}     {0}{upmath}{16}
    \NewMathSymbol{\upartial}{0}{upmath}{40}
    \NewMathSymbol{\leqslant}{3}{AMSa}{36}
    \NewMathSymbol{\geqslant}{3}{AMSa}{3E}

     \let\le=\leqslant
     \let\ge=\geqslant
  \else
    \def\umu{\mu}
    \def\upi{\pi}
    \def\upartial{\partial}
  \fi
\fi


\pageoffset{-2.5pc}{0pc}

\loadboldmathnames



\pagerange{1--7}    
\pubyear{1989}
\volume{226}

\begintopmatter  

\title{Towards Multiple-Star Population Synthesis}
\author{P. P. Eggleton$^1$}
\affiliation{$^1$Lawrence Livermore National Laboratory, 7000 East Ave, Livermore, CA94551, USA}

\shortauthor{P. P. Eggleton}
\shorttitle{Multiple Systems}


\acceptedline{Accepted .... . Received ....}

\abstract {The multiplicities of stars, and some other properties, were 
collected recently by Eggleton \& Tokovinin, for the set of 4559 stars with 
Hipparcos magnitude brighter than 6.0 (4558 excluding the Sun). In this 
paper I give a numerical recipe for constructing, by a Monte Carlo technique, 
a theoretical ensemble of multiple stars that resembles 
the observed sample. Only multiplicities up to 8 are allowed; the observed set
contains only multiplicities up to 7. In addition, recipes are suggested for 
dealing with the selection effects and observational uncertainties that attend 
the determination of multiplicity. These recipes imply, for example, that to
achieve the observed average multiplicity of 1.53, it would be necessary to
suppose that the real population has an average multiplicity slightly over 2.0.
\par   This numerical model may be useful for (a) comparison with the results 
of star and star cluster formation theory, (b) population synthesis that does 
not ignore multiplicity above 2, and (c) initial conditions for dynamical 
cluster simulations.}

\keywords{stars: statistics; stars: general}
\maketitle  

\section{Introduction}
\par Recently Eggleton \& Tokovinin (2008; ET08) counted the stellar systems 
with multiplicity from 1 to 7 in the set of stars brighter than mag. 6 on the
Hipparcos scale. There are 4558 such sytems in total (excluding the Sun), with 
2716, 1438, 285, 86, 20, 11 and 2 having
multiplicity 1 to 7. We refer to this as the Hipparcos-Bright Multiple System 
Catalogue (HBMSC). It is, of course, provisional: a few new components are
found every year. The present paper proposes a simple model distribution
that will generate the observed population, including the multiplicities.
Selection effects are modeled, so that allowance can be made for the
fact that some potential visual binaries or sub-binaries may not be resolvable 
(depending partly on distance), and some potential spectroscopic binaries
may be undetectable because of unfavorable luminosity or mass ratios, or
broad lines as in many OB stars. I estimate that to obtain the observed
multiplicity fraction of 1.53 (ET08), i.e. an average of 1.53 components per 
system, I need an actual multiplicity fraction of about 2.0 or more. I 
attempt to fit the distribution of periods (and sub-periods), as well as of 
multiplicity and period.

\par Statistics on multiplicity are useful for at least three reasons.
\sep (a) They are a statement about the final product of star formation, and
thus something of a check on theoretical models of this difficult process.
\sep (b) There are some stellar evolutionary scenarios that are importantly
modified by, or even wholly dependent on, the existence of triple companions;
thus we need to know something of the probability of triple (or higher
multiple) companions.
\sep (c) They represent a starting point for N-body dynamical simulations of
clusters, and more generally for any population-synthesis project.
\pn Note however that the HBMSC can provide statisics only for systems
with mass above about $1\Msun$, since very few systems of lower mass are
included among the bright stars. Although many low-mass multiples have been
recognised, there does not exist a comparable wealth of data on, say, the
5000 {\it nearest} stars.

\par In relation to point (a), it can be questioned how representative are
the systems in the general Galactic field of what is produced directly in
Star Formation Regions (SFRs). This is not the place to attempt a detailed
discussion of star formation, but there seems no doubt that the great
majority of stars form in localised, relatively dense, SFRs, and are then
released or scattered into the Galactic field. Since only about 1\% of stars
are currently in clusters, it follows that the field stars are the great
bulk of what is produced in SFRs, and so it is not unreasonable to see them
as representative of those products. Some of the systems have probably
been modified, by N-body encounters, during both the formation process
and the later escape process. One might attempt to distinguish between the
formation of {\it stars} and the formation of {\it systems}, but although a
semantic distinction might be made I doubt if any clear {\it physical} 
distinction can be made. The widest systems that can be found in
the field are usually much too wide to have had a chance to form in the
relatively compact SFRs where they were born, but presumably formed during
the process of evaporation that put them into the field. Those field
systems that are not very wide can reasonably be supposed to be much the same
as when they formed in the SFR. Numerical attempts to model star formation in
SFRs will presumably have to address the N-body interactions and the 
dissipation into the field as well as the gas-condensation stage, and so it 
is not unreasonable that they should be constrained to some extent by the 
statistics of multiplicity in field stars.

\par These remarks also relate to point (c). The distribution of multiples
in a  cluster will certainly differ from that of the field, by missing the
widest field multiples. But if one starts an N-body cluster with the
multiplicity distribution suggested below, the widest systems will be
very quickly broken up. It is not unreasonable to suppose that the closer
systems, which do not get broken up quickly, might be representative of
what should be expected in a young cluster.

\par In Section~2 I describe the Monte Carlo procedure I use, applied to a 
model in which stellar evolution, but only so far of single stars, is taken 
from a table of evolved 
stellar models. In Section~3 the procedure is generalised from single 
to multiple stars; a procedure called {\tt Create} constructs a theoretical 
population of multiple systems. Although the issue of {\it interactive} 
binary and triple systems is briefly discussed, it is left for 
future work. In Section~4 I describe a `theoretical observatory', a 
procedure called {\tt Observe}, which looks at the multiple systems
generated by {\tt Create} and decides which systems and subsystems 
should be recognisable as visual or spectroscopic or eclipsing
binaries, and which not. Thus a `raw' model from {\tt Create} is 
turned into a model that can in principle be compared with what is 
actually seen, in the HBMSC. In Section~5 a comparison is made, and I discuss 
some applications of the model to the problems mentioned in the second 
paragraph.

\par I would like to emphasise that I am not claiming to find a definitive
model of multiplicity, but only a reasonable model. For instance, a
definitive model would need many more mass ratios than are currently 
well-known, and in particular rather extreme mass ratios, such as are very
difficult to determine reliably at present. In addition, there is an
obvious degeneracy between the {\tt Create} stage and the {\tt Observe}
stage, since one can imagine that many more binaries (and higher multiples)
are created that have parameters that are hard to observe, and so are
excluded at the {\tt Observe} stage.  

\section{Method}

\par We would like to model the statistics with some set of (cumulative)
distribution
functions. Suppose for the sake of argument that all stars are single stars. 
Each star is determined by two parameters, mass ($m$) and age ($t$). I ignore 
metalicity for the time 
being -- and in fact throughout this paper -- although it would not be 
difficult in principle to include it. For determining which stars are visible
above some apparent magnitude, a third parameter, distance ($d$), has also to 
be assigned. The cumulative distributions
of these parameters will be some functions of the three variables.  There is no
{\it a priori} reason to suppose separability, i.e. that the distribution is
the product of three elementary distributions, each in one variable only.
We select three random numbers $X,Y$ and $Z$, uniform in the interval $(0,1)$,
and then generate $m$ from some global (cumulative) IMF, $m=m(X)$, 
$t$ from an age distribution $t=t(Y,m)$, and finally
$d$ from a distance distribution $d=d(Z,m,t)$. 
The order is immaterial, although of course the actual
functional forms will depend on the order selected. The distribution
over distance will have to fall off at large distances fast enough to avoid
Olbers' paradox; the formulation given below -- equation (10) or its inverse
equations (12) and (13) -- is roughly 
spherical within $\ts100\thn$pc, roughly disk-like to $\ts 1000\thn$pc and
becomes negligible beyond $\ts 3\thn$kpc.
\par The number of stars brighter than a star of luminosity $L_0$ at distance
$d_0$, i.e. that have
$${L\o d^2}\tge {L_0\o d_o^2}\ ,\eqno(1) $$
out of a total population of $N_*$ stars is an integral 
$$N=N_*\int_{\rm unit\ cube}H\{L(X,Y)-L_0{d^2(Z)\o d_0^2}\}\thn\thn dX\thn 
dY\thn dZ\ ,\eqno(2)$$
where $H$ is the Heaviside step function. The integral can be rewritten as
$$N\=N_*\int_{\rm unit\ square}\thn Z_v(X,Y)\thn dX\thn dY\ ,\eqno(3)$$
where $Z_v$ is the fraction of stars at distance less than $d_v$, and $d_v$
is the distance out to
which a particular star as a funtion of $X, Y$ is visible. This distance $d_v$
is given by
$$d_v^2\= d_0^2\thn{L(X,Y)\o L_0}\  .\eqno(4)$$

\par Given that stars are discrete rather than continuous, we replace the 
integral by a Monte Carlo approximation. Suppose we select
$n^2$ points randomly in the $X,Y$ unit square. Then we approximate the 
integral (3) by
$$N\={N_*\o n^2}\thn\sum_{i=1}^{n^2} \thn Z_v(X_i,Y_i)\ .\eqno(5)$$
It is clear that only stars that come from a rather small fraction of the
$X,Y$ plane will populate the night sky visibly to the naked eye: we know
from experience that only stars more massive than about $0.8\Msun$, i.e.
$\tls 10\%$ of stars from a reasonable IMF, can be bright enough, and the 
most massive and brightest stars must be very young, e.g. age less than 
$\ts 10^7\thn$yrs at masses $\tgs 10\Msun$. 
We can expect that a rather small fraction of the $n^2$ systems will 
contribute at all, but some of these will contribute rather heavily.
We therefore use `importance sampling', i.e. a transformation that maps the 
unit square in $X,Y$ from the unit square in say $\Xp,\Yp$ so that the 
region near $X=1, Y=0$, where most {\it visible} stars come from, is 
sampled more densely. Following Eggleton et al. (1989; EFT89) I use
$$        X=1-(1-\Xp)^2\ ,\w Y=1-(1-\Yp)^{(1-\Xp)}\ .\eqno(6)$$ 
In equation (5) we
have to insert an extra factor, the Jacobian of the transformation:
$$N\={N_*\o n^2}\thn\sum_{i=1}^{n^2}\thn Z_v(X_i,Y_i)\thn
{\pa(X,Y)\o\pa(\Xp,\Yp)}\ .\eqno(7)$$

\par Equation (7) gives a fractional number of stars at each sampling point.
We quantise this in the following way. At the $i$th element the value 
$N_*Z_vJ/n^2$ in equation (7), where $J$ is the Jacobian, is some fractional
number, say $k+f$, where k is an integer and $0\tle f\tl 1$. Then select a 
further random number, uniform in $[0,1)$. If this is less than $f$ then 
$k+1$ `cloned' stars are allowed, but if greater, then $k$ clones.
A value for $n$ of 200 was found to give a reasonable compromise between
speed and detail.


\par The formula
$$m\=0.3\thn\left({X\o 1-X}\right)^{0.85}\ ,\eqno(8)$$
gives a distribution which, in terms of $\log m$, is peaked at $X=0.5,$ 
and falls off with a moderately Salpeter-like slope of $2.23$ for 
$m\tgs 0.8\Msun$. The median mass, at $X=0.5$, is $m=0.3\Msun$.
For age we use the simplest distribution
$$t=Y,\eqno(9)$$
$t$ being the age in units of $10^{10}$yrs. Although for a total Galactic
population such a uniform rate of star production is unlikely to be
realistic, most of the bright stars are less than a tenth of the Galaxy's
age, and a uniform birth rate is not so unreasonable in this span.

\par Formulae (8) and (9) can be seen as `plucked out of thin air'.
However a distribution like (8) has the desirable quality of
resembling the Salpeter distribution at large masses, but turning
over in a reasonable way at low masses, below $0.3\Msun$. The value
of the turnover mass and of the slope below that mass are not
determinable from the HBMSC, because the latter includes hardly
any stars below $\ts 1\Msun$. The slope at the high-mass end
{\it is} in principle determinable, say by doing some least-squares
fitting; but my initial value of 0.75 seemed to produce too many
high-mass relative to intermediate-mass stars, and a second guess
at 0.85 seemed to be about right, as quantified by a $\chi^2$ test
given near the end of this Section. The assumption in (9) of a reasonably
uniform star formation rate over the last Gyr is similarly bound to be
incorrect in detail (i.e. on a time scale of Myrs, and a lengthscale
of tens of parsecs), and yet is a fairly reasonable model on longer 
timescales and lengthscales.

\par The cumulative distribution $Z(d)$ is approximated by
$${1\o Z}={h_0h_1^2\o d^3}+{h_1^2\o d^2}+1\ ,\eqno(10)$$
so that
$$\ \ \ Z\propto d^3\w{\rm if}\w d\tls h_0\tll h_1$$
$$\w\propto d^2\w{\rm if}\w h_0\tls d\tls h_1$$
$$\w\ts 1\w{\rm if}\w d\tgs h_1\ .\eqno(11)$$
We need the inverse of this, $d(Z)$. If
$$ x\eq{3h_0\o 2h_1}\sqrt{{3(1-Z)\o Z}}\ ,\eqno(12)$$
then 
$$ d\thn\={3h_0\o x}\thn\cos({1\o3}\cos^{-1}x)\w{\rm if}\w x\tle 1$$
$$ \ \ \={3h_0\o x}\thn\cosh({1\o3}\cosh^{-1}x)\w{\rm if}\w x\tge 1\ .
\eqno(13)$$

\par A refinement that we make to this $d(Z)$ model is that we
take $h_0$, which is essentially the scale height of the disc, to be
dependent on age:
$$h_0(t)\=ht^{0.3}\ ,\ h\= 200\ ,\eqno(14)$$
with $h,h_0$ in parsecs and $t$ in units of $10^{10}\thn$yrs, as before.
The constant $h_1$ is taken as $1\thn$kpc.
Another refinement is that reddening, at 1 mag. per kpc, is included. 

\par Once again formulae and coefficients are apparently plucked out of 
thin air,
but are based nevertheless on a general picture of the Solar neighbourhood
that I believe is widely accepted: that the distribution is roughly
spherical for old stars but more disc-like for young stars. I have simply
adopted what I believe is about the simplest formulation that describes
this. I make no attempt to improve the fit by varying some coefficients.

\centerline{\psfig{figure=HRD.ps,height=5.0in,bbllx=0pt,bblly=0pt,bburx=560pt,bbury=780pt,clip=}}
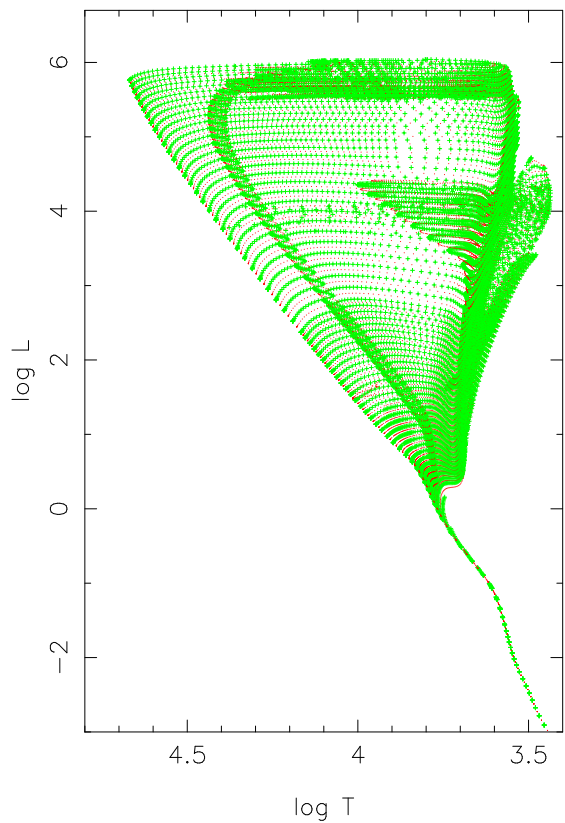
{Fig 1 -- Hertzsprung-Russell diagram for 171 stellar models with 
initial masses in the range $\sc 0.1 - 63\Msun$. Evolution was terminated 
when it became rapid, approaching a final compact remnant; except that 
an upper age limit of $\sc 10\thn$Gyr was also imposed.}
\halfline

\par I used my stellar-evolution code (Eggleton 1971, 1972, Eggleton et al. 
1973, Eggleton 2006; Pols et al. 1995) to evolve 171 single-star models 
with (log) masses of
$$\log\thn m\= -1.0\ (0.02)\ 0.0\ (0.01)\ 0.6\ (0.02)\ 1.80\ .\eqno(15)$$ 
All had a roughly solar composition: $X=0.70,\thn Y=0.28,\thn Z=0.02.$
They were evolved from the Zero-Age Main Sequence (ZAMS) either until
$10^{10}$yrs old or until fairly near the end of their lives.
For the lowest masses this meant very little evolution in practice. For
intermediate masses I stopped the evolution either at the very rapid
phase of evolution towards the white-dwarf region, or shortly after the 
onset of carbon burning in the core. A rate of mass loss by stellar wind
was adopted (Eggleton 2006), so that intermediate-mass stars less massive
originally than $4.2\Msun$ would become white dwarfs less massive than
$1.05\Msun$. All estimates of mass-loss rates for red supergiants are
very tentative; the one used here gave relatively little mass loss in the
(initial) mass range $6-20\Msun$, but proportionately much increased mass 
loss at still higher masses.

\par For every tenth timestep of each evolutionary sequence, the following
six numbers were stored: $ t, m(t), \log L, \log R,$ $ \log R_{\rm max},$ 
and $I_{\rm ev}$. $R_{\rm max}$ is the maximal radius in the evolution
prior to age $t$, useful for knowing whether binary interaction with a fairly
close companion will have occurred, and $I_{\rm ev}$ is an integer that
describes crudely the evolutionary state, from 1 for an MS star to 10 for a
neutron star or black hole. Fig. 1 shows a theoretical Hertzsprung-Russell 
diagram (HRD) for this collection of stars. To avoid some interpolation 
which would often be of a highly non-linear nature, when the Monte Carlo 
procedure of equation (8) selects an initial 
mass I replace it by the nearest value in the sequence (15).
\def\z{\ \ \ \ \ \ \ \ }
\begintable*{1}
\caption{{\bf Table~1. Distribution over spectral type.} }
\halign{%
\hfil\rm#&\hfil\rm#&\hfil\rm#&\hfil\rm#&\hfil\rm#&\hfil\rm#&\hfil\rm#
&\hfil\rm#&\hfil\rm#&\hfil\rm#\cr
\noalign{\vskip 12pt}
         &     O &   eB &   lB &      A &     F &     G &    K  &     M  &   tot.\cr
observed &\z 38&\z 401&\z 653&\z   928&\z  578&\z  587&\z 1042&\z  331&\z  4558\cr
&&&&&&&&\cr
single &    97&   620&   897&   886&   468&   492&   881&   225&  4565\cr
rms    &    16&    44&    46&    65&    25&    28&    79&    87&   138\cr
&&&&&&&&\cr
OB down&    83&   440&   929&   915&   491&   509&   923&   260&  4550\cr
rms    &    18&    42&    60&    41&    42&    24&    95&   102&   123\cr
&&&&&&&&\cr
multiple&   47&   402&   699&   917&   581&   529&  1041&   335&  4550\cr
rms    &     9&    45&    55&    39&    34&    35&   101&    56&   148\cr
}
\tabletext{eB stands for early B (B0 -- B3.5), and lB for later B.
\pn `Single' refers to the single-star model of Section 2.
\pn `OB down' is a model, described at the end of Section 2, where stars 
over $6\Msun$ and nearer than $250\thn$pc are discounted.
\pn `Multiple' is the same as `OB down', but with multiplicities up to 8 
allowed as in Section 3.}
\endtable

\par Table~1 presents some results. The distributions (8) for mass, (9) for 
age and (10) for distance were used, and only $N_*$, the total population,
was varied  to get about the right number (to much better than 1\%) of visible 
stars. The distribution over spectral type (Table~1, row
labeled `single'), to some extent equivalent to the distribution over mass at 
least for types F and earlier, is not very satisfactory compared to the 
observed distribution. The ratio computed/observed is nearly 3 for O stars, 
and ranges between 1.4 and 0.7 for the remaining spectral types.

\par Although we might try to ameliorate this by playing with the IMF, it 
is not difficult to see that much of the disagreement is really a 
consequence of the fact
that the distributions of mass, age and distance are not independent of each
other, as is implicit in equations (8) -- (10); although equation (10) for
the distance dependence does include an age dependent term via equation (14).
The regions that form
massive stars are very clumpy, with several hundred parsecs between them,
and the Sun appears to be located within a `bubble' (Cox \& Reynolds 1987) of 
100 -- 250$\thn$pc
in which no formation of {\it massive} stars has taken place recently,
i.e. within about 75$\thn$Myr, the lifetime of a $6\Msun$ star. O stars
have formed much more recently than that within say the Orion SFR (Star
Forming Region) at about $450\thn$pc, and are easily above $H_p=6$, but 
there is no O star nearer than about 250$\thn$pc. Thus one way to improve
the agreement at O and early B is to exclude stars above $6\Msun$ that are
nearer than $250\thn$pc, and that was used in the line of Table~5 labeled 
`OB down'. This is equivalent to saying that the mass, age and distance
distributions are not in fact disjoint; but it is numerically easiest
simply to omit systems that have mass above a threshold and distance
below another threshold.

\par We might try some more alterations, particularly of the IMF, to improve 
agreement further, but in practice much of the remaining disagreement 
disappears when we include multiple systems, in the next Section. The row 
labeled `multiple' in Table~1 is essentially the same model as the `OB down' 
model, but with multiplicity included as will be discussed in the next Section.
The agreement is either slightly or in some cases considerably better. The 
overall improvement is at least partly because 
massive systems, as we will see shortly, appear to be more highly multiple than
less massive systems, and so some O and eB systems are broken down to later 
types. A $\chi^2$ test comparing the four distributions over spectral type
listed in Table 1 gives, of course, very poor agreement between the
observed distribution and either of the first two theoretical distributions,
`single' and `OB down'. But we get a $\chi^2$ of 11.2 when comparing the
`multiple' with the `observed' distribution, which is acceptable at the 10\% 
level. No doubt this could be improved by tweaking some coefficients, but
this does not seem necessary.

\par For each of the `single', `OB down' and `multiple' lines in Table 1, ten 
Monte Carlo models were run differing only in the random numbers generated.
What is listed is the average of these ten, and the rms variation. A few
preliminary runs were done in order to estimate the value of the normalisation
constant $N_*$ of equation (7) that would give a final count (averaged over 
the ten runs) of approximately the number 4558 observed.

\section{\bf A Model of Multiplicity}

\begintable*{2}
\caption{{\bf Table~2.}  {\bf Bifurcation Probability $U_0$ as a Function of 
Mass and Hierarchical Level.}}
\halign{
#\hfil&\hfil #&\hfil #&\hfil #&\hfil #&\hfil #&\hfil #&\hfil#&\hfil#&\hfil#\cr 
&&&&&&&&&\cr
\ \ $m=$&\w 0&\w .01&\w .09&\w .32 &\w 1 &\w 3.2 &\w 11 &\w  32 &\w$\infty$\cr
\ \ $j=$& 1 &\w 2   & 3     &4      &5      &6      &7      &8      &9      \cr
$l=0$&\w0.40&\w 0.40&\w 0.40&\w 0.40&\w 0.50&\w 0.75&\w 0.88&\w 0.94&\w 0.96\cr
\w\ 1&\w0.18&\w 0.18&\w 0.18&\w 0.18&\w 0.18&\w 0.18&\w 0.20&\w 0.60&\w 0.80\cr
\w\ 2&\w0.00&\w 0.00&\w 0.00&\w 0.00&\w 0.00&\w 0.20&\w 0.33&\w 0.82&\w 0.90\cr
\w\ 3&\w0.00&\w 0.00&\w 0.00&\w 0.00&\w 0.00&\w 0.00&\w 0.00&\w 0.00&\w 0.00\cr
}
\tabletext{The values for $m\tls 1$ are notional, because very few systems of 
this low a total mass appear in the HBMSC. However they are reasonably 
consistent with what is known about the {\it nearest} 66 systems.}
\endtable

\par When deciding on the multiplicity of a computed stellar system, I continue
to use random numbers, thinking in terms of a process of successive 
bifurcation; although for practical reasons I allow no more than 3 successive 
bifurcations so that only multiplicity up to octuple is considered.
Note however that this is {\it not} a claim that real multiples are
formed by successive fragmetation during a pre-main-sequence contraction phase.
It is simply an acknowledgment of the well-established fact that the great
majority of multiple systems are highly hierarchical (Evans 1968), and
such systems by definition consist of binaries nested inside wider binaries.
A rather small proportion of systems is non-hierarchical, in the sense that
3 or more components are found at roughly equal distances from each other.
Seventeen such systems were identified in the HBMSC of ET08, but these are 
among the
least certain. Their inclusion or exclusion will make little difference to
the following.

\par We have already selected a mass $m$, which we now interpret as the
{\it total} mass, and the age $t$. When considering whether to bifurcate or 
not, we choose two more random numbers uniformly probable in (0,1), $U$ and 
$V$ say, to determine (i) the mass ratio $Q$ (with a finite probability of 
zero, i.e. no bifurcation)  and (ii)
orbital period $P$. Of course, both $P$ and $Q$ are irrelevant if the 
choice of $U$ is found to imply no bifurcation after all. $P$ and $Q$ might, 
indeed probably will, depend on $m$ as well as $U, V$, and also on the 
hierarchical depth $l$, i.e. on whether we are creating a binary,
a sub-binary or a sub-sub-binary. We allow only $l$ = 0, 1 or 2. At each 
potential bifurcation we choose a fresh $U$ and $V$, so that we may need as 
many as 7 $U,V$ pairs, although with some probability that there may be no 
bifurcation at all, and only one $U$ is used. 

\par First we determine whether the current hierarchical level $l$ bifurcates 
or not. I specify (Table~2) a tabular function 
$U_0(m,l)$, $l$ being the hierarchical depth. Then $U\tge U_0$ implies no 
bifurcation, and $U\tl U_0$ implies bifurcation with a mass ratio that we 
can suppose depends 
in some manner on $(U_0-U)/U_0$, a ratio which is always between zero and 
unity, and with uniform probability. Actually the entries in Table~2 are 
treated as functions of a variable $j$ related to total mass $m$ by
$$j\ =\ 5+2\log\left({1+100m\over 100+m}\right)\z.\eqno(16)$$
The variable $j$ ranges from 1 to 9 as $m$ ranges from zero to infinity,
and I interpolate linearly in the Table for non-integer values of $j$.

\par Then for $l=0$, and assuming that the random $U$ does imply bifurcation, 
a new random number $V$ is selected in (0,1). I take the period to be
$$P_{l=0}\=10^5{V^2\o (1-V)^{2.5}}\ ,\eqno(17)$$
in days. This expression gives a distribution of $\log P$ 
peaking at about $10^5$d, dropping rather slowly towards large P and also 
dropping, a little more rapidly, towards small $P$. The lowest ten percentile
is at $\ts 1.3.10^3\thn$d, and the highest at $2.5.10^7\thn$d. For $l\tg 0$, 
I adopt
$$P_l\=0.2\thn P_{l-1}\thn 10^{-5V}\ .\eqno(18)$$ 
The median period ratio is $10^{-3.2}$. These determinations of outer or 
inner periods may well be expected to depend also on all previous choices, 
in particular on $m$ and $l$. They might depend on $t$ as well,
if we are talking about a cluster in which orbits can soften or harden as
time advances; but for the time being I do not allow such complexity.

\par The mass ratio $Q$ can be determined from the previous $U$ and 
corresponding bifurcation probability $U_0$ by, for instance,
$$Q\=\left({U_0-U \o U_0}\right)^{\alpha}\ ,\ \ \alpha=f(m,l,P_l),\eqno(19)$$
whether $l$ is 0, 1 or 2. Values of $Q$ less than 0.01 are replaced by 0.01.
Because, as noted above, $(U_0-U)/U_0$ is uniformly distributed in (0,1),
distribution (19) is just a power law. Obviously with enough information it 
will be better to have a more complicated distribution, but several attemped 
determinations of the mass-ratio distribution have been presented as power laws. 
For example, Kouwenhoven (2006) obtains, for near-infrared companions to lB/A 
stars in the Sco~OB2 association, a differential distribution $\propto Q^{-0.33}$.
This corresponds to $\alpha \ts 1.5$ in equation (19). Tokovinin (2000),
following Lucy \& Ricco (1979), finds a marked preference for `twins', i.e
near-equal-mass pairs, at fairly short periods ($\tls 25\thn$d). This would 
require a low power like $\alpha\ts 0.15$ if one wanted 50\% of
systems to have $Q\tg 0.9$. For the present I assume, following EFT89, that 
$\alpha=0.8$ for all periods above $25\thn$d, and for 75\% of those with shorter
periods, but for the remaining 25\% of the latter I take
$$Q=0.9+0.09\thn\left({U_0-U \o U_0}\right)\ .\eqno(20)$$

\par One would like to do the same analysis for the 4500 
{\it nearest} stars, but although information on these stars, particularly on 
L and T dwarfs, has increased by leaps and bounds over the last decade, there 
is nothing like as complete a study of their multiplicity as already exists 
for the $\ts4500$ {\it brightest} stars. However, for the 66 nearest stars, 
including the Sun, multiplicities 1, 2, 3 and 4 have frequency
42, 16, 8 and 0. These are roughly the values we expect using the values for
$U_0(m,l)$ in the columns of Table~2 appropriate to $m\tls 1$.

\par Obviously any bifurcation will have to be reversed, i.e. ignored, if the
period selected turns out to be too small for the components to exist 
separately. This may be seen to beg the question of contact binaries. I have 
argued elsewhere (Eggleton 1976, 1996) that contact binaries are not `born' 
in contact, 
but their periods evolve towards contact as a result of one or both of the
following processes: magnetic braking and Kozai cycles, both assisted by
tidal friction. I return to this point
in Section~5. The question is slightly academic in the present context, as 
there are only three contact binaries in the HBMSC; but mergers may mean that
there are some {\it former} contact binaries as well. However it is evidently 
necessary to place some lower limit on $P$ based on the sum of the radii of
two ZAMS stars relative to their separation. It is also necessary to 
have some upper limit. I take the upper limit to be $10^{10}$d ($27\thn$Myr) 
for hierarchical depth $l=0$, and, according to equation (19), 
$0.2P_{l-1}$ for hierarchical depth 
$l\tg 0$. The former comes loosely from the consideration that wider orbits 
would be disrupted by independent neighbouring stars, and the latter from the 
dynamical stability of coplanar triples (Harrington 1975).
All orbits are assumed circular for the present. It would be easy to 
randomise the eccentricities if their cumulative distribution is postulated;
but this would introduce the issue of modeling tidal friction to vary the
eccentricity with time, and for the present I want to adhere to simplicity.
The lower limit for $P_l$ comes, as mentioned before, from the condition that
a star should not fill its Roche lobe on the ZAMS.

\par A set of systems formed by this procedure will be called a `Raw Theoretical 
Bright Multiple Star Catalogue' or RTBMSC; in the next Section I describe the 
`Theoretical Observatory' which decides which systems, subsystems and 
subsubsystems should be recognisable, and which not. The result of that process
will be a Theoretically Observed Bright Multiple Star Catalogue, or TOBMSC. 
Some systems of particularly high luminosity appear several times over in 
the RTBMSC, according to the `cloning' recipe between equations (7) and (8).
{\tt Create} assigns a random inclination to each pair, at each hierarchical 
level, and does this independently for each system, including each clone. 
Such random inclinations should in fact influence the dynamical stability, but
I do not include such an effect yet. For the present, they influence only
the likelihood that a subsystem will be detectable as either a spectroscopic
or eclipsing binary by the Theoretical Observatory (Section~4).

\par Equations (17) to (20) are simpler than might have been expected. We
could expect that the probability distribution of period might depend on the
mass, and that similarly the distribution of mass ratio might depend on mass 
{\it and} the period. My prescription (19), (20) for $Q$ is period-dependent, 
as described in the text, but the effect of this period dependence is fairly
modest in view of the modest numbers that have $P\tl 25\thn$d.

\par The computer code {\tt Create} contained in principle 45 adjustable
constants, although only 12 of the 27 constants in Table~2 (corresponding to
the larger elements in the rows $l=0,\thn 1$ and 2) were in practice adjusted. 
In  addition, only 6 of the 18 remaining constants were varied, from guessed 
initial values, to try and improve the agreement. The 12 not varied include, 
for instance, almost all of the coefficients and exponents in equations 
(8) -- (14) and (17) -- (20). In most cases it was possible to estimate from 
an eyeball analysis of the HBMSC what values would be plausible, and these 
values did indeed appear to work without further tweaking. It would be only 
a slight exaggeration to say that the coefficient and two exponents of
equation (17), for example, were `plucked out of thin air'. But they were
based on an eyeball analysis of the HBMSC section of Table 6, along with
the suspicion that the rather marked shortage of systems there with periods
of 1 -- $100\thn$yr at early spectral types might be an effect of observational 
selection, as is indeed found in the next Section.

\begintable*{3}
\caption{{\bf Table~3.} {\bf Velocity and Temperature Thresholds as functions 
of Spectral Type.}}
\halign{
\hfil#&\hfil#&\hfil #&\hfil #&\hfil #&\hfil #&\hfil #&\hfil #&\hfil #\cr 
&&&&&&&&\cr
Type            \z&\z     O&\z  eB&\z  lB &\z A &\z  F&\z  G &\z \z K&\z  M \cr
$\vcr$(km/s) \z&\z  30.0&\z  25.0&\z   20.0&\z 12.0&\z  6.0&\z   2.0&\z  2.0&\z  3.0 \cr
$T_{\rm eff}(K)$\z&\z 31000&\z 15500&\z  10200&\z 7400&\z 5900&\z  4880&\z 4000&\z      \cr
}  
\tabletext{The first row of numbers is the minimal radial velocity amplitude
considered measurable at that spectral type. The second row is the minimal 
effective temperature considered belonging to the spectral type. Note that the 
G/K temperature boundary (4880K) has to be appropriate to giants rather than 
to dwarfs.}
\endtable

\par Once an $n$-tuple system has been created, the various components can be 
evolved, all according to the same prescription as described in Section~2. I 
do not (yet) put in a binary interaction, such as Roche-lobe overflow (RLOF). 
It is feasible to put in back-of-the-envelope prescriptions (Hurley et al. 
2002), but I prefer, in a future version, to utilise the capacity of
my stellar evolution code to follow RLOF in some circumstances (Nelson \&
Eggleton 2001, de Mink, Pols \& Hilditch 2007). But a substantial proportion 
of potentially interacting systems are quite difficult to follow numerically, 
because the RLOF can become very rapid (Paczy{\'n}ski \& Sienkiewicz 1972). In
addition winds, including those linked magnetically to a component and capable
of draining angular momentum from the component, and via tidal friction from
the binary, can affect the orbit. Such winds are included in my binary
evolution code, which can work in a mode where both components, {\it and} the 
orbital period, and in addition the eccentricity, which is also modified by 
tidal friction, can be solved for in a single implicit time step. This process 
even extends to models in which both stars overfill their Roche lobes, and 
influence each other's structure via internal luminosity transfer as is 
frequently hypothesised for contact binaries (Lucy 1968, 1976, Yakut \& 
Eggleton 2005); but this mechanism still contains some very uncertain physics, 
and it is also unusually expensive to compute since evolution progesses in 
cycles on a thermal timescale, while lasting for something like a nuclear 
timescale. Thus the number of timesteps is larger than usual by a factor of 
several hundred.
\par The worst cases of rapid mass transfer, usually hypothesised to lead to 
common-envelope evolution (Paczy{\'n}ski 1976), and to either very close white 
dwarf + MS star binaries or else to a merger, are not dealt with yet, even 
by the most sophisticated version of my evolution code. In practice, these 
are likely to be the majority of RLOF cases, and so for the present I simply 
annotate those systems in which one component is, or has been at some time in 
the past, larger than its Roche lobe.

\begintable*{4}
\caption{{\bf Table~4.} {\bf Multiplicity Frequency.}}
\halign{%
\rm#\hfil&\qquad\rm\hfil#&\qquad\rm\hfil#&\qquad\rm\hfil
#&\qquad\rm\hfil#&\qquad\rm\hfil#&\qquad\rm\hfil#
&\qquad\rm\hfil#&\qquad\rm\hfil#&\qquad\rm\hfil#&\qquad\rm\hfil#\cr
Sample&$n=1$&2&3&4&5&6&7&8&tot.&av.\cr
\noalign{\vskip 10pt}
HBMSC &   2716&   1438&    285&     86&     20&     11&      2&      0&   4558&   1.53\cr
&&&&&&&&&&\cr
RTBMSC&   1459&   2179&    517&    202&    101&     44&     30&     18&   4550&    2.04\cr
rms   &     63&    111&     37&     28&     10&     11&     13&      7&    148&    0.03\cr
&&&&&&&&&&\cr
TOBMSC&   2649&   1472&    302&     85&     29&      9&      3&      0&   4550&    1.55\cr
rms   &     96&     85&     37&     17&      5&      3&      3&      0&    148&    0.02\cr
}
\tabletext{The last column is the average multiplicity or `companion
star fraction (CSF)', defined as $\sum_n nN_n/\sum_n N_n$. Ten models were
generated by {\tt Create}, averaged to give the line RTBMSC, then theoretically
observed by {\tt Observe}, and finally averaged to give the line TOBMSC.}
\endtable

\par Quite a few massive systems will have produced supernovae by now, even
though one or two OB stars (and even AF stars) remain in the system. This is 
partly why we need multiplicity significantly higher in O stars than in much 
later stars. Although we expect that such neutron stars should have ejected 
themselves from their natal systems by means of asymmetric mass ejection 
(Shklovskii 1970, Lyne \& Lorimer 1994, Hansen \& Phinney 1997), I choose in 
the present work to keep them in the system so that, for example, I can 
estimate from the RTBMSC how many neutron stars should have been produced by 
the systems of the HBMSC: $220\pm 24$. Many more systems will have produced white 
dwarfs, and these are rather more likely to have remained in their original 
systems, although some, perhaps the majority, may have merged with a companion 
star. The number of white dwarf companions, seen or not, and 
merged or not, is $371\pm 60$ in the RTBMSC. Only 20 are known in the HBMSC,
which suggests that many WD companions have not so far been recognised as such.

\section{\bf The Theoretical Observatory}

\par I use the code described in the previous Section and called {\tt Create}
to create a `raw' catalogue of theoretical bright multiple systems, the RTBMSC. 
Then I use a code called {\tt Observe} which reads in the
output catalogue from {\tt Create} and decides which components would actually
be identifiable, either as visual systems, spectroscopic systems or eclipsing
systems. This works on very simple principles, outlined below, and converts
the kind of statistics seen in the middle sections of Tables 4, 5 and 6 into those
of the bottom sections. Let us call this the `Theoretically Observed BMSC' or
TOBMSC.

\par For visual binaries to be resolvable we require that one or other of
$$\Delta\hp\tl 2.5(\rho-0.1\asec)\ ,\eqno(21)$$
$$\Delta\hp\tl 4.5\thn\log(\rho/0.05\asec)\ ,\eqno(22)$$
be satisfied, where $\rho$ is the separation in arcsecs and $\Delta\hp$ is the 
modulus of the difference in Hipparcos magnitudes. A further condition is that 
the fainter (sub)component should be brighter than 14th magnitude. Criterion 
(21) is roughly appropriate to normal visual measurements, and the much stronger
criterion (22) is roughly appropriate for high-resolution measurements with
speckle or adaptive optics (Sterzik \& Tokovinin 2002). For the present I assume 
simply that 50\% of stars have been examined by one process and 50\% by the 
other, chosen at random.

\par For a spectroscopic binary (SB) to be identifiable I assume that the 
radial velocity amplitude of the brighter component, $K_1$, is above a 
threshold value $\vcr$ that is given in Table~3, and further that the
period is less than $\ts 50\thn$yr:
$$K_1\tg \vcr\ \ ,\ \ P\tl 18000\thn{\rm d}\ .\eqno(23)$$ 
The threshold $\vcr$ depends on spectral type. Theoretical SBs can be single or
double lined, depending on the luminosity ratio. The value of $K_1$ includes an
inclination which was assigned randomly to each sub-binary in the {\tt Create}
stage. This implies the assumption that inclinations at different hierarchical
depths are uncorrelated. It is hard to test this: Muterspaugh et al. (2006) 
showed that only six triples have unambiguous determinations of the inclination
of one orbit to the other (as distinct from the inclination of each orbit to 
the line of sight). It is unlikely that 6 data points would give a clear 
determination of the statistics. In fact the inclinations were not very 
consistent with a complete lack of correlation, nor with complete correlation.
Sterzik \& Tokovinin (2002) considered 135 visual triples, and estimated
the mean angle between orbits as $67\deg\pm 9\deg$. This value would be $90\deg$
if the two angular momenta were uncorrelated; thus there appears to be some
correlation but not a great deal. However, for most of these systems there
was a $180\deg$ ambiguity in the position angle of one or other line of nodes,
which creates an ambiguity in the relative angle of the angular momenta; this
ambiguity was missing in the much smaller sample of Muterspaugh et al. (2006). 

\par The inclination assigned in {\tt Create} also says, along with the
period and radii, whether a system or subsystem will eclipse, and further gives
an estimate of the depth of eclipse. Provided the eclipse depth is sufficient, 
such eclipsers are added to the tally of recognised theoretical binaries. 
However all theoretical eclipsers turn out to be theoretical SBs, whether 
single or double-lined. I do not at present count ellipsoidal variables, 
although they are included in the HBMSC.
\begintable*{5}
\caption{{\bf Table~5.} {\bf Multiplicity Frequency by Spectral Type.}}
\halign{\rm#\hfil&\hfil #&\hfil #&\hfil #&\hfil #
&\hfil #&\hfil #&\hfil #&\hfil #&\hfil #&\hfil #\cr
sp&1&2&3&4&5&6&7&8&tot.&av.\cr
HBMSC&&&&&&&&&&\cr
  O & \z  13& \z  10& \z   7& \z   4& \z   2& \z   2& \z   0& \z   0&  \z 38& \z2.42\cr
 eB &    213&    125&     46&      8&      6&      1&      2&      0&    401&   1.70\cr
 lB &    353&    215&     61&     21&      1&      2&      0&      0&    653&   1.63\cr
  A &    501&    336&     63&     17&      6&      5&      0&      0&    928&   1.61\cr
  F &    304&    219&     36&     15&      4&      0&      0&      0&    578&   1.61\cr
  G &    323&    217&     33&     12&      1&      1&      0&      0&    587&   1.56\cr
  K &    739&    263&     32&      8&      0&      0&      0&      0&   1042&   1.34\cr
  M &    270&     53&      7&      1&      0&      0&      0&      0&    331&   1.21\cr
tot.&   2716&   1438&    285&     86&     20&     11&      2&      0&   4558&   1.53\cr
RTBMSC&&&&&&&&\cr
  O&      7&     15&      6&      7&      6&      2&      2&      2&     47&    3.31\cr
 eB&     76&    162&     56&     39&     38&     15&     10&      6&    402&    2.80\cr
 lB&    247&    339&     71&     25&      9&      4&      3&      2&    699&    1.92\cr
  A&    320&    453&    102&     27&      9&      3&      3&      1&    917&    1.88\cr
  F&    211&    264&     67&     26&      6&      4&      2&      1&    581&    1.93\cr
  G&    172&    269&     58&     19&      7&      3&      1&      1&    529&    1.94\cr
  K&    351&    490&    122&     45&     17&      7&      6&      4&   1041&    1.99\cr
  M&     77&    188&     36&     15&      9&      6&      4&      1&    335&    2.19\cr
tot.&  1459&   2179&    517&    202&    101&     44&     30&     18&   4550&    2.04\cr
TOBMSC&&&&&&&&&\cr
  O&     14&     15&      8&      6&      2&      1&      0&      0&     47&    2.41\cr
 eB&    192&    126&     44&     24&     12&      3&      1&      0&    402&    1.88\cr
 lB&    458&    205&     28&      6&      2&      0&      0&      0&    699&    1.41\cr
  A&    561&    300&     47&      9&      1&      0&      0&      0&    917&    1.46\cr
  F&    333&    193&     41&     11&      3&      1&      0&      0&    581&    1.56\cr
  G&    278&    205&     36&      8&      1&      0&      0&      0&    529&    1.58\cr
  K&    616&    325&     74&     16&      6&      2&      1&      0&   1041&    1.54\cr
  M&    198&    102&     23&      6&      3&      2&      1&      0&    335&    1.56\cr
tot.&  2649&   1472&    302&     85&     29&      9&      3&      0&   4550&    1.55\cr
}
\tabletext{Early B stars (B0 -- B3.5) are called `eB'; later B stars are 
called `lB'. Wolf-Rayet stars are included under O; S and C stars under M. 
The second last column is the total number; the last column is the average 
multiplicity.
\pn HBMSC: the observed Hipparcos-Bright Multiple Star Catalogue (ET08).
\pn RTBMSC: a Raw Theoretical Bright Multiple Star Catalogue (Section~3).
\pn TOBMSC: a Theoretically Observed Bright Multiple Star Catalogue (Section~4).
\pn Each row of RTBMSC and TOBMSC is an average of 10 simulations, rounded
to the nearest integer. Consequently the row or column for `total' is not
necessarily the {\it exact} total.
}
\endtable

\def\y{\ \ \ \ }
\begintable*{6}
\caption{{\bf Table~6.} {\bf Period Distribution in Systems and Subsystems.}}
\halign{\rm#\hfil&\hfil #&\hfil #&\hfil #&\hfil #
&\hfil#&\hfil#&\hfil#&\hfil#&\hfil #&\hfil #&\hfil #&\hfil #&\hfil #&\hfil #\cr
$\log P$(yr)&&-3.0\y&-2.0\y&-1.0\y&0.0\y&1.0\y&2.0\y&3.0\y&4.0\y
&5.0\y&6.0\y&7.0\y&8.0\y&tot.\cr
HBMSC&&&&&&&&&\cr
  O &   0&   5&  11&   5&   0&   4&   3&   6&  12&   1&   0&   0&   0&   47\cr
 eB &   0&  25&  41&  21&  14&  21&  26&  29&  33&  21&   3&   0&   1&\y  235\cr
 lB &   0&  25&  53&  24&  20&  43&  62&  63&  48&  16&   8&   0&   0&  362\cr
  A &   0&  27&  62&  25&  46&  78&  78&  66&  47&  24&   9&   0&   0&  462\cr
  F &   0&  14&  33&  24&  39&  46&  61&  44&  26&  14&   3&   1&   0&  305\cr
  G &   1&   7&   9&  20&  40&  38&  49&  38&  32&  23&   6&   1&   0&  264\cr
  K &   0&   4&   4&  11&  56&  35&  35&  42&  35&  26&   4&   0&   0&  252\cr
  M &   1&   0&   0&   4&   9&  10&   4&   9&   9&   3&   3&   0&   0&   52\cr
tot.&   2& 107& 213& 134& 224& 275& 318& 297& 242& 128&  36&   2&   1& 1979\cr
RTBMSC&&&&&&&&\cr
  O&    0&   16&   21&   12&   11&   15&   12&   11&    5&    2&    1&    1&    0&  107\cr
 eB&    1&   84&   96&   87&   89&  101&   99&   77&   48&   25&   10&    6&    1&  724\cr
 lB&    5&   29&   42&   53&   79&  106&  129&  101&   57&   27&    9&    4&    0&  641\cr
  A&    8&   34&   52&   62&   97&  142&  159&  128&   77&   31&   12&    9&    0&  810\cr
  F&    8&   29&   38&   51&   64&   86&  104&   74&   50&   21&    8&    7&    0&  539\cr
  G&    5&   19&   33&   48&   58&   92&   87&   78&   47&   19&    7&    5&    0&  497\cr
  K&    4&   50&   77&   86&  122&  166&  174&  164&   94&   56&   25&   13&    1& 1032\cr
  M&    1&   18&   35&   35&   36&   56&   48&   72&   60&   25&    6&    7&    0&  399\cr
tot.&  32&  278&  394&  434&  556&  764&  811&  705&  439&  206&   77&   52&    2& 4749\cr
TOBMSC&&&&&&&&&\cr
  O&    0&    8&   10&    6&    2&    2&    6&    7&    5&    2&    1&    1&    0&   49\cr
 eB&    0&   42&   56&   28&    8&    7&   35&   47&   36&   16&    7&    5&    0&  286\cr
 lB&    0&   15&   21&   17&    7&   18&   51&   66&   38&   20&    7&    3&    0&  263\cr
  A&    0&   18&   29&   36&   27&   33&   79&   85&   56&   22&   10&    7&    0&  399\cr
  F&    1&   18&   27&   36&   36&   30&   52&   52&   33&   12&    5&    5&    0&  307\cr
  G&    1&   13&   21&   34&   42&   59&   33&   44&   29&   12&    5&    3&    0&  297\cr
  K&    0&   37&   59&   57&   95&   93&   35&   69&   45&   30&   11&    5&    0&  536\cr
  M&    0&   12&   25&   19&   17&   28&    6&   24&   35&    6&    4&    2&    0&  178\cr
tot.&   2&  164&  249&  232&  233&  270&  297&  394&  275&  120&   48&   31&    1& 2315\cr
}
\tabletext{HBMSC: the observed Hipparcos-Bright Multiple Star Catalogue (ET08).
\pn RTBMSC: a Raw Theoretical Bright Multiple Star Catalogue (Section~3).
\pn TOBMSC: a Theoretically Observed Bright Multiple Star Catalogue (Section~4).
\pn Each row of RTBMSC and TOBMSC is an average of 10 simulations, rounded
to the nearest integer. Consequently the row or column for `total' is not
necessarily the {\it exact} total.
}
\endtable

\section{\bf Discussion}

\par Tables 4, 5 and 6 present some details of the multiplicity model that appears to 
represent reasonably well the observed data of the HBMSC. Table~4 shows the 
distribution over multiplicity. The TOBMSC has somewhat too many triples and
quintuples, somewhat too few binaries and sextuples, and roughly the right 
average. The last, of course, is 
not by accident; the entries in Table~2 were varied from ones initially 
guessed until the average came out about right. I should emphasise again that 
only those values in Table~2 that relate to $m$ ({\it total} mass) above unity 
are tested by the HBMSC. 

\par The top third of Tables 5 and 6 gives the HBMSC data from ET08. The 
middle third gives the equivalent data that emerges from the {\tt Create} 
stage described in Section~3. The bottom third gives the result of processing 
the middle third through the selection effects of the {\tt Observe} stage of 
Section~4. I ran ten cases differing only in the random number selection. 
Tables 4 -- 6 present the average of these ten, and in some cases the rms 
scatter. The 10 cases gave an average total of $4550\pm 148$ systems. 
Because the middle and lower thirds are both averages of 10 cases, the 
rows and columns labelled `total' in Tables~5 and 6 are not necessarily the
{\it exact} totals but only the approximate totals of the integers in the
bodies of these Tables. Note 
that the RTBMSC in Table~4 produced $18\pm 7$ octuple systems, but none of 
these fully survived the scrutiny of the Theoretical Observatory.

\par When in Table~5 we compare the distribution over spectral type 
(second-last column) of the TOBMSC with the HBMSC we get a $\chi^2$ of 11.2, 
as already quoted in Section 2. When we compare the distribution over 
multiplicity (last row) of the 
TOBMSC with the HBMSC we get a $\chi^2$ of 7.3. The first value is on the margin
of significant or non-significant discrepancy at the 10\% level, while the
second favours no significant discrepancy. However a $\chi^2$ test is not
really going to validate (or contradict) the model, since we could have the
same $\chi^2$ using either (a) a model of multiplicity that gives more 
multiples, combined with a model of selection effects that makes them harder
to recognise, or (b) the converse of that. I am not claiming that both halves
of the model are correct, but only that they represent a reasonable starting 
point. Most stellar population synthesis calculations discount entirely
multiplicity greater than two, and I believe the present model is somewhat 
better than that.

\par The distribution over spectral type (second-last column of Table~5) is 
the same as in the last two rows (`multiple') in Table~1, apart from the fact 
that  Table~1 contains the rms spread as well as the average. The only 
difference between the model in Table~5 and the `OB down' model of Table~1 is 
that multiples are now allowed.  Probably the remaining discrepancies can be 
improved by varying the IMF a little. However there are at least two other ways
of altering them: (i) `convective overshooting', an effect modeled 
in my stellar-evolution code on the lines of Schr{\"o}der et al. (1997) and 
Pols et al. (1997), can vary the length of time spent on the MS, differentially
as a function of mass; and (ii) small changes, of order 50K, in the temperature
boundaries of spectral types in Table~3 can send quite a lot of systems from
say type A to type F. I prefer for the time being the rather considerable 
simplicity of the model used here, without trying to further improve the 
agreement.

\par The selection effects implemented in {\tt Observe} are very crude, but
they seem to imply that in order to get approximately the observed average 
multiplicity of 2.42 for O stars, we need to start with an average well in excess 
of 3.0. This average has to drop fairly rapidly to about 2 at later spectral 
types in order that the observed multiplicity should drop to $\ts 1.6$, and 
even lower for M stars. The average multiplicity in the entire ensemble, before
the {\tt Observe} stage, was $2.04\pm 0.03$. After {\tt Observe}, it dropped to
$1.55\pm0.02$. In the HBMSC, and of course also the TOBMSC, triples and higher 
multiples constitute about 9\% of systems, but in the RTBMSC they constitute
20\%, which I suggest is a more realistic assessment of their frequency. 

\par In Table~6, the `observed' periods of the HBMSC that are fairly long, in 
particular
those over $\ts 200\thn$yr, are estimates that are based on Kepler's Law, a 
measured angular separation and parallax, and a mass based on spectral type.
They are obviously very uncertain, but perhaps by not much worse than one
bin (a factor of 10) either way. We see that the effect of selection, for 
OB stars, is to turn a largely unimodal period distribution as seen in the
RTBMSC section of Table~6  into a strongly bimodal distribution as seen in 
the TOBMSC and HBMSC sections. The `missing' systems at periods of 
$\ts 1 - 100\thn$yr are presumably too wide (and broad-lined) for radial 
velocity curves, and too close for visual resolution.

\par In fact a slight bimodality can be seen even in the RTBMSC period 
distribution
for type O, which comes from the fact the the shorter periods are more probably
from inner pairs via equation (18) than from outer pairs via equation (17).
But the bimodality is much more marked when the selection effects are allowed
for, and runs all the way from type O to type A, with a minimum at about
$10-100\thn$yrs as observed.

\par The most obvious discrepancy between the HBMSC and TOBMSC sections of
Table~6 is in the bottom left-hand corner, for types G/K/M and periods
$\tls 1\thn$yr. The fact that there are many in the TOBMSC and and few in the 
HBMSC is due to binary interaction. I have emphasised that this is not (yet)
included in the {\tt Create} code, and so the model is comfortable producing
say a $0.01\thn$yr binary containing a well-evolved red giant, that ought to 
have filled its Roche lobe some time ago. Such a system might now resemble
$\alpha$ Leo (Gies et al. 2008) where a late B subgiant has a low-mass
companion that is (probably) a white dwarf, in a $40.1\thn$d circular orbit. 
This orbit
was not known previously, and is not included in ET08. 204$\pm 27$ such 
systems can be expected. Some will probably have merged into single stars,
and others to some other part of the Table. 17 binaries or sub-binaries are
known in the HBMSC to be semidetached, i.e. currently undergoing RLOF. A 
further handful, such as $\phi$ Per, $\gamma$ Cas, $\theta$ Tuc and AY Cet, 
and now $\alpha$ Leo, are believed to be post-RLOF
objects. But we can be confident that there are many more in this category
that would be hard to recognise by any means. 

\par Leaving aside the bottom left-hand corner, we can see that there is 
reasonable agreement over the rest of the Table. I do not believe it would be 
helpful to try and quantify this by some $\chi^2$ statistic, or to overegg
the pudding by trying to make the agreement even better with adjustments
to, for instance, the period distributions (17) and (18).

\par There appears to be another discrepancy: the total number of orbits in
the HBMSC is 1979, and in the TOBMSC is 2315. But this is because there are
several systems or subsystems (in fact 384) in the HBMSC that do not have a 
measured, or even estimated, period. These are for example the `astrometric
accelerators' of Makarov \& Kaplan (2005), radial velocity variables where
probably the motion is orbital but no orbit has been determined 
(Nordstr{\"o}m \& Andersen 1985) and about 40 Ba stars for which
a white-dwarf companion is expected and yet no orbit has yet been measured.

\par One of the more important uncertainties in the model is the mass-ratio 
distribution (19). It is unlikely that a single value of the exponent $\alpha$
will apply independently of $m, P$ and hierarchical depth $l$. It would be 
desirable to check
this much more rigorously, by using observed mass ratios. However in 
observational analysis the mass ratio is one of the more difficult quantities
to measure. A spectroscopic binary has to be double-lined, with the secondary
preferably not very faint so that its amplitude can be well determined. This
means that mostly only systems with $Q$ fairly close to unity are measurable,
and so the regime of $Q\tls 0.5$ is not well tested. The TOBMSC model of
Tables 5 and 6 produced $198\pm 21$ double-lined and $854\pm 56$ single-lined 
spectroscopic systems or subsystems. In the HBMSC 638 systems or subsystems 
are SB1 or SB2. The discrepancy is fairly large, but is quite probably because 
all (theoretical)
systems or subsystems with $P\tls 50\thn$yrs and large enough radial velocity 
amplitude (Table~3) are recognised as SB, whereas it is likely that many real 
systems in the period range of say $10 - 50\thn$yrs have not been recognised.
I am not claiming that a global choice of $\alpha=0.8$ is {\it correct}, but
only that it works reasonably well to produce an ensemble like that observed.

\par The TOBMSC contained $85\pm 12$ eclipsing binaries, against 130 in the 
HBMSC. The RTBMSC (Table~6) contained as many as 32 very short-period 
($P\tl 0.001\thn$yr) binaries, most of which were eclipsing; but these were 
all double-M-dwarf {\it sub}-binaries of systems whose primary had the type 
listed, and all but two on average were undetectable in the Theoretical 
Observatory. The HBMSC contains
two systems at such short periods, both being subsystems in fact. The 
well-known M-dwarf eclipsing sub-binary in the sextuple $\alpha$~Gem is almost 
in the same category, but its slightly longer period puts it in the 
$0.001-0.01\thn$yr bin.

\par Barium-rich (Ba) stars are thought (McClure 1983, Boffin \&Jorissen 1988)
to be binaries where the primary is now a white dwarf, and where the secondary,
a G/K giant, has been contaminated from s-processed material in the envelope of
the white dwarf's precursor when it was near the tip of the Asymptotic Giant 
Branch (AGB). There are 53 Ba stars in the HBMSC: 11 of them have known SB
orbits, with periods ranging from 285 to 6500 days; and 3 of them have 
{\it known} white dwarf companions, notwithstanding the difficulty of 
recognising a white dwarf companion whose luminosity will usually be one part
in $10^3$ or $10^4$ of the red giant. In the TOBMSC, all systems containg a 
white dwarf, a G/K giant, and with period in the range $100 - 10000\thn$d were 
identified as potential Ba stars. There were 37$\pm 15$, as compared with the 
53 in the HBMSC. But many Ba stars in the HBMSC are only mildly enriched in Ba,
and if this can be achieved in wider binaries with periods up to $20000\thn$d
then there is no discrepancy ($56\pm 25$). 

\par I emphasise again that the HBMSC only includes systems more massive than
about $1\Msun$ in total. Coincidentally, but conveniently, this is also
the set of stars that are capable of evolving significantly in the course
of the Galaxy's lifetime. Thus this model may be quite adequate for exploring
the evolution of an ensemble of multiples (including by definition singles) on 
the scale of a galaxy, or a galactic cluster. In the context of a cluster, we
can be reasonably sure that the wider systems will be broken up quite quickly,
provided that the cluster remains rich for a reasonably long time. However,
current thinking is that {\it all} stars form in rich or fairly rich SFRs. 
This must therefore include the rather highly-multiple, yet
also quite wide, O star multiples of the HBMSC, which may indeed be the result 
of (a) gravitational diffusion inwards of the most massive stars of a cluster, 
and (b) dissipation of the cluster because of the ejection of significant mass
by, for instance, supernova explosions. I believe that one of the more
rigorous tests of star-formation models will be that they produce the right
spectrum of multiplicity, as well as the right spectrum of masses and
orbital periods.

\par A striking result of rather recent analyses (Tokovinin et al. 2006, 
Pribulla \&Rucinski 2006), confirmed by ET08, is that short-period binaries 
($P\tl 3\thn$d) are very commonly in triples, and arguably are exclusively in
triples (or higher multiples); and this then implies that
tripleness is {\it necessary} for the production of a short-period binary.
The necessary mechanism may be the combination of Kozai cycles and tidal
friction: KCTF (Mazeh \& Shaham 1979, Kiseleva et al. 1998, Eggleton \& 
Kisseleva-Eggleton 2006, Fabrycky \& Tremaine 2007). It may 
be that the process of star formation does not {\it directly} produce binaries 
with periods of less than say $0.1 - 1\thn$yr, but it produces enough triples, 
with suitably oblique orbits, that these short periods can be generated by
the KCTF mechanism, in the course of $10^3 - 10^9\thn$yr depending on the
initial parameters. Once an orbital period has been reduced to $\ts 2 - 3\thn$d
by KCTF, those systems which contain late-type dwarfs (F/G/K/M) may evolve
to still shorter periods by the quite different process of magnetic braking
and (again) tidal friction, or MBTF. This can reduce the period to the point
where a contact binary is formed, on a timescale that may be roughly
$10^7 - 10^{10}\thn$yr.

\par The reader will note that there is something of a contradiction between
\sep (a) offering a formulation that is intended to give roughly the right
distribution of periods at age zero,
\sep (b) generating arbitrarily short periods via equations (17) and (18),
although with the distribution truncated when the period is so short that the
stars would be touching on the ZAMS, and
\sep (c) hypothesising that the shortest periods ($\tls 3\thn$d) are in fact 
due to the KCTF and MBTF mechanisms, and not to {\it formation} at such short
periods.
\pn I would hope, in a future analysis, to investigate whether by considering
the interactive evolution of triples hypothesis (c) can be supported, but until
then it seems desirable to produce about the right frequency of such
short-period systems {\it ab initio}.

\par Part of the above apparent contradiction is because `age zero' is rather
hard to define. Returning to the brief discussion of star formation in the
introduction, the bright stars in the Solar neighbourhood presumably formed
in SFRs, and then escaped as the SFRs evaporated; but they may have been
subject to dynamical interaction within their parent SFRs for at least
several million years, so that their multiplicities and periods may have been
modified in that interval. `Age zero' can be surprisingly well defined for
individual stars when their nuclear evolution alone is being discussed. But 
`age zero' is much harder to define when binaries and higher multiples are 
being considered.

\par The purpose of this paper has been to produce a model for the multiplicity
statistics of an observed complete magnitude-limited sample of substantial
size. Such a model must 
include selection effects, but even the best such model will not be definitive.
In particular, one can add to almost every observed system a T dwarf that would
hardly be recognisable by current technology, if located at a suitable distance
from the primary component. We can only attempt to find something like a lower
limit to multiplicity; but I believe that the model proposed here is something
like a lower limit.

\section*{Acknowledgments}
This work was performed under the auspices of the U.S. Department of 
Energy by Lawrence Livermore National Laboratory under Contract 
DE-AC52-07NA27344. I gratefully acknowledge the help
of the Centre des Donn{\'e}es Stellaires (Strasbourg), and of the
Astronomical Data System.

\bye